\documentclass[11pt]{article}
\usepackage{geometry}                
\usepackage{graphicx}
\usepackage{amssymb}
\usepackage{epstopdf}
\usepackage{caption}
\usepackage{threeparttable}
\usepackage{array}

\usepackage[colorlinks = true,
            linkcolor = blue,
            urlcolor  = blue,
            citecolor = blue,
            anchorcolor = blue,
            hypertexnames=false] 
            {hyperref}

\def\authorlist#1#2{
    \vskip 0.4in
\begin{center}\begin{large} {\bf  #1 } \end{large}
    \vskip 0.2in
              #2
     \vskip 0.2in
   \end{center}
}

\begin{document}

\title{\textbf{Snowmass 2021 \\ Underground Facilities for the Cosmic Frontier \\ Topical Report}}

\maketitle

\authorlist{J.~Cooley$^*$, S.~Hertel, H.~Lippincott$^*$, K.~Ni, E.~Pantic \newline \textit{$^*$Liaisons from other Snowmass Frontiers.}}{}

\tableofcontents





\section{Direct detection of particle-like dark matter }


Detecting dark matter directly using sensitive underground detectors is one of the most promising ways to address the question of what is dark matter. During the last decade, noble liquid detectors probed and excluded large parameter spaces for heavy weakly interacting massive particle (WIMP) as the candidate for dark matter and new technologies are being developed to probe light dark matter with extremely low energy thresholds. Experiments have searches for not only conventional nuclear recoils from dark matter nucleus scatterings, but also from electronic recoils resulting from dark matter electron scattering and absorption from other dark matter candidates such as the axion-like particles. 

The next generation underground experiments will search for dark matter in two directions. One is to use larger target mass detectors to further probe smaller cross sections in the WIMP parameter space and the other is to probe lower mass dark matter which requires new technologies with extremely low threshold. Both directions will eventually be limited by astrophysical neutrino background. Once dark matter is detected, experiments with directional sensitivities and experiments with longer exposure and larger target masses are needed for the time-dependent signatures, such as annual modulations, to verify their cosmic origin and to further study their properties. 







\section{Future experiments and their need}
\label{futureDMexperiments}

The next generation underground dark matter search experiments fall into two main detector classes: large noble liquids (liquid argon and liquid xenon) detectors with the primary goal to probe lower WIMP cross sections and low-threshold detectors (cryogenic bolometer experiments and other techniques) to probe light dark matter candidates. Their needs for underground facilities are described separately below. We also comment briefly on the need of other technologies for dark matter searches. 

\subsection{Noble liquids}

The next generation noble liquid experiments will use 50 to 300 tons of liquid xenon or liquid \textit{underground} argon (argon extracted from underground sources) to probe heavy WIMPs into the ``neutrino fog". Future Projects at the planning stage (from CF1 report) include:
\begin{itemize}
\item The PandaX-xT experiment at CJPL using $\sim$50 ton liquid xenon. 
\item A 50--100 ton liquid xenon observatory by the XLZD consortium, location undetermined 
\item A 300 ton ARGO liquid \textit{underground} argon observatory at SNOLAB 
\end{itemize}

There are also smaller scale noble liquid experiments at the conceptual stage, such as the DarkSide-LowMass program using about 1.5-ton liquid argon. 

Maintaining the large amount of cryogenic liquid in a stable cold liquid and its supporting systems, such as distillation columns, requires a significant continuous cooling power, which can result in heat-loads to the underground facility which must be carefully considered. Liquid nitrogen cooling can be supplied in addition to industrial cryogenic cooling. For example, the XENON1T infrastructure features a 10 m$^3$ LN$_2$ tank underground, serving several subsystems. A ten times larger LN$_2$ tank supplying the need for the next generation liquid xenon will be needed.

Many of the challenges in the coming noble liquid projects are due to the scale and volume of the target material itself, and the even larger volumes of surrounding active veto and shielding.  The detector must be in an experiment hall that is large both in floor space and ceiling height.  Large noble liquid experiments also require additional underground staging space for `cylinder farms' or other storage of large amounts of Xe or Ar at room temperature, while waiting for the cosmogenic activated isotopes to decay away. At a shallow depth, cosmogenic activation of certain isotopes, such as Xe-137, will be a concern if the liquid xenon target will be used to search for $^{136}$Xe $0\nu\beta\beta$ signals. Argon in the atmosphere contains cosmogenically produced radioactive isotopes, mainly $^{39}$Ar but also $^{37}$Ar or $^{42}$Ar which can be a significant background. The concentration of these thee isotopes is greatly reduced in argon extracted from underground source, but the production of cosmogenic radionuclides after extraction must be minimized. Typically, $^{37}$Ar activation in liquid xenon or liquid argon at the surface will decay away in several months of storage at underground. Further reduction of $^{37}$Ar in xenon can be achieved using cryogenic distillation column as used in XENON1T/nT. 

The future noble liquid dark matter experiments, both LAr and LXe, would benefit greatly from on-site or near-site isotopic separation ability, to mitigate for example a small air leak or cosmogenically-activated radioisotopes at surface, introducing a trace amount of noble radioisotopes (e.g. $^{\mathrm{37}}$Ar,$^{\mathrm{39}}$Ar,or $^{\mathrm{85}}$Kr), and to reduce the radioactive radon emanating from the detector inner surfaces. This can be achieved using on-site cryogenic distillation columns, such as use in XENON1T~\cite{U2-Aprile1} and XENONnT. The XENON1T/nT distillation column, requires 5.5\,m of height, and releases some 10s of kW of heat into the cavern. The XENONnT radon distillation column~\cite{U2-Murra:2022mlr} is 3.8\,m in height and requires a few kW of cooling power. Both the height and heat load requirement of a distillation column should be considered when siting future large noble liquid DM experiments.

Muon-induced neutron background should be reduced to a negligible level combining the deep underground rock overburden and sufficiently-thick active shield. LZ and XENON1T/nT~\cite{U2-XENON:2017lvq} liquid xenon detectors use water, while DarkSide-20k liquid argon detector uses liquid argon as the shield for external and muon-induced neutrons. Further neutron background from the detector material is reduced by using liquid scintillator based veto detectors with demonstrated high efficiency for neutron tagging. Since not all the underground facilities are allowing usage of liquid scintillator, gadolinium doped in the water, or a liquid argon veto coupled with gadolinium loaded acrylic are new technologies  employed.  These veto's neutron tagging efficiency will need to be demonstrated with the current generation experiments. Safe handling of these shielding liquids (gadolinium doped liquid scintillator or water) shall be taken into consideration in regard of the environmental concerns.

Assembling the large multiple-stage detector is a challenge in terms of cleanliness (better than ISO-6), low-dust and low-radon requirement (better than 100~mBq/m$^3$). The current approach of transporting the inner detector assembled in an above-ground clean room will not be feasible due to the large future detector sizes. Large underground low-radon clean room with staging areas is needed to assemble the inner detector parts. Sufficient vertical space and cranes are needed to lift heavy detector parts during assembling. The underground facilities will also need to balance the needs of continuing experiments when performing any construction works such as new excavations. 

Finally, the safety of handling the huge cryogen target, for both human safety (asphyxiation etc.) and enormous expense of the target gas itself, needs to be carefully taken into account during the design and operation of the experiments.

\textbf{Related: Large Bubble Chambers}

Target liquid at pressure and in a superheated metastable state, they are the 500-kg to 5-ton scale PICO-500 and PICO-5T experiments, and a 1-ton scale liquid argon Scintillation Bubble Chamber (SBC).  These experiments are similar to standard noble liquid experiments in facility requirements, in that pressure vessel and cryogenic safety topics require host lab input and interaction. Underground laboratories should provide recommendations on how to handle these safety topics in next generation experiments.

\subsection{Cryogenic bolometers}

Many future experiments will focus on DM nuclear recoils of sub-keV energy, with a signal either largely or entirely in the phonon or heat channel.  Such experiments rely on mK temperatures for this sensitivity, achieved via 3He/4He dilution refrigeration.  The 3He/4He dilution refrigeration technology enforces a space constraint on such experiments, particularly in the vertical direction.  Even if the target masses themselves are small (e.g., $<$1~kg) the experiment as a whole requires significant vertical space for the opening and closing of the fridge and the shielding external to the cryostat.  A vertical space of at least 4m is highly beneficial.  It is possible to configure a dilution-refrigerator-based experiment to require less vertical space, but at the cost of increased complexity and decreased cooling power.

The vibration environment is a second item of concern for many cryogenic bolometer technologies.  The target mass of such experiments are held in some fashion to the support structure, and a slip in this holding, even a slip at the atomic scale, can induce a visible `dark rate' of signal into the phonon system.  While multiple groups are investigating mitigation methods either through alternative target holding methods or vibrational isolation~\cite{U2-Maisonobe}, the environment itself must also be considered.  A typical environmental goal may be to keep these vibrations below 10$^{-7} \mathrm{g\sqrt{Hz}}$ at all frequencies.

Many bolometric sensors require a quiet electromagnetic environment as well.  Superconducting electronics (e.g. SQUID amplifiers) are highly sensitive to noise across a wide range of frequencies, and are also sensitive to a DC magnetic field.  Faraday cage mitigations and cold filtering can help, but the E\&M environment should be kept in mind when selecting underground sites, developing underground laboratory space, and evaluation of how experiments can share a facility without cross-interference.

As the threshold of bolometric technologies continues to be pushed to lower and lower energies, the reduction of backgrounds specific to sub-keV energies is now an active, complementary research area.  At these low energies, backgrounds include Cherenkov or luminescence backgrounds from insulators near the detectors~\cite{U2-Du}, and also heat-only events which are still mysterious in origin~\cite{U2-Adari}.  Given that very low-threshold experiments are currently dominated by non-cosmogenic backgrounds, it is difficult to assess the depth requirement of future low-threshold experiments until that R\&D is more mature.

However, while depth requirements for \emph{full experiments} with cryogenic bolometers are difficult to assess, the need for underground space for \emph{R\&D associated with cryogenic systems} is important for continued development success. Specifically, the traditional strategy is to complete R\&D above ground, then deploy underground. In current and future generations of cryogenic bolometer development, low-rate, `heat-only events' from instrumental effects are better constrained in the quiet and cosmic-ray shielded environments of underground laboratories. Stated another way, cryogenic experiments with long duration events suffer a challenge of characterization due to event pile-up in above ground test systems. This is a recognition of the value of underground space availability for developmental and R\&D-scale efforts in advancing cryogenic bolometer dark matter searches.

\subsection{Other technologies}

As described in the CF1 report, there are a multitude of other dark matter detection technologies, including skipper CCD (OSCURA), Point-contact Ge (CDEX-100/1T), Low-pressure gas detectors (directional detectors, CYGNUS), Scintillator (SABRE, COSINE-200, others), and Superheated water (SNOWBALL).   At their current stages of development, each of these technologies, except CDEX, is (for now) relatively compact in footprint, requiring underground space at a level similar to or smaller than the cryogenic bolometer class.  Also, each of these technologies is relatively robust to lab environmental factors such as vibration, etc. or would otherwise perform well if the needs of the above noble liquid and cryogenic bolometer experiments are met.

\section{Conclusions}

Unraveling the properties of dark matter will continue being pursued through direct detection at deep underground laboratories with larger targets (noble liquids) and new experiments being planned. The facilities hosting the current generation experiments are mostly full. There is a clear need for expanded underground space, tailored to the needs beyond late 2020s. 

These new underground space should specifically include: large spaces for large experiments, specially noble liquids (liquid argon and liquid xenon) and small spaces for smaller experiments (cryogenic bolometer \& ‘other technologies’); large radon-free clean rooms for detector assembly and installation for the next generation experiments. In addition, large staging areas are needed to store low-background detector materials, such as xenon or underground argon gases, and utilities, such as cryogenic facilities.







\end{document}